\documentclass[prl,twocolumn,aps,showpacs]{revtex4}
\usepackage{graphics,graphicx,epsfig}
\usepackage{color}

\begin{document}
\title{Universal fluctuations in the growth of semiconductor thin films}
\author{R. A. L. Almeida${}^{1,(a)}$, S. O. Ferreira${}^{1,(b)}$, T. J. Oliveira${}^{1,(c)}$ and F. D. A. Aar\~ao Reis${}^{2,(d)}$
\footnote{a) Email address: renan.lisboaufv@gmail.com \\
b) Email address: sukarno@ufv.br \\
c) Email address: tiago@ufv.br (Corresponding author) \\
d) Email address: reis@if.uff.br }}
\affiliation{${}^{1}$ Departamento de F\'isica, Universidade Federal de Vi\c cosa, 36570-000, Vi\c cosa, MG, Brazil\\
${}^{2}$ Instituto de F\'\i sica, Universidade Federal Fluminense, Avenida Litor\^anea s/n, 24210-340 Niter\'oi RJ, Brazil\\
}

\date{\today}

\begin{abstract}
Scaling of surface fluctuations of polycrystalline CdTe/Si(100) films grown by hot wall epitaxy are
studied. The growth exponent of surface roughness and the dynamic exponent of the auto-correlation
function in the mound growth regime agree with the values of the Kardar-Parisi-Zhang (KPZ) class.
The scaled distributions of heights, local roughness,
and extremal heights show remarkable collapse with those of the KPZ class,
giving the first experimental observation of KPZ distributions in $2+1$ dimensions.
Deviations from KPZ values in the long-time estimates of dynamic and roughness exponents
are explained by spurious effects of multi-peaked coalescing mounds and by effects of grain shapes.
Thus, this scheme for investigating universality classes of growing films advances over the simple
comparison of scaling exponents. 
\end{abstract}

\maketitle

Non-equilibrium interface dynamics and kinetic roughening theories are fascinating topics in statistical mechanics
due to the emergence of scaling invariance and universality and a wide number of applications \cite{barabasi}.
A paradigm in this field is the Kardar-Parisi-Zhang (KPZ) equation \cite{KPZ}
\begin{equation}
 \frac{\partial h (x,t)}{\partial t} = \nu \nabla^{2} h + \frac{\lambda}{2} (\nabla h)^{2} + \eta(x,t),
\label{eqKPZ}
\end{equation}
which is a hydrodynamic approach to an interface driven by white noise ($\eta$) and subject to linear and nonlinear
(slope-dependent) tension mechanisms. A number of applications of the KPZ equation was suggested in the last decades
based on the comparison of scaling exponents of surface roughness \cite{barabasi}. An increased interest in
this problem was observed after the exact calculation of the height distribution (HD) of the $1+1$-dimensional
KPZ equation in the growth regime \cite{SasaSpo1} and its connection to random matrix theory \cite{TW1}
(long after the first calculation of the HD of some lattice models in the KPZ class \cite{johansson}).
These results were confirmed by numerical works \cite{SidTiaSil1} and in experiments on fluid
convection and colloidal particle deposition \cite{TakeSano,yunker}.
In $2+1$ dimensions, which is the most interesting case for applications, recent numerical studies of
lattice models \cite{healy,SidTiaSil3} showed that the KPZ HD have the same scaling properties
of $1+1$ dimensions, with two universal distributions for flat and radial growth.
However, the exact solution and the experimental confirmation are still lacking.

Thin film and multilayers form the basis of modern micro and optoelectronic industry \cite{ohring}.
The application of kinetic roughening theory to describe their morphology helps to understand
their physical properties and to predict the conditions for growth of novel structures.
Here, cadmium teluride thin films are grown on $Si(001)$ substrates by hot-wall epitaxy and their
surface morphology is studied along those lines. $CdTe$ is a direct gap semiconductor that, in the 
form of either bulk crystal, thin films or quantum dots, has widespread use as solar cells, X-ray 
and $\gamma$ detectors, and other optoelectronic devices \cite{cdteappl}. 

A thorough atomic force microscopy (AFM) study of grown $CdTe$ films provides very accurate distributions of
heights, local roughness,
and extremal heights, which, together with global roughness scaling, provide the first experimental confirmation
of those KPZ distributions in $2+1$ dimensions. Local roughness exponents are consistently related to the shapes
of surface mounds. The nontrivial mound coalescence and coarsening dominated by grains in (111)-direction
explains the excess velocity that leads to KPZ scaling. This establishes a robust procedure to understand local
and global features of a film surface from existing theories.

\begin{figure*}[ht]
\includegraphics[width=17.50cm]{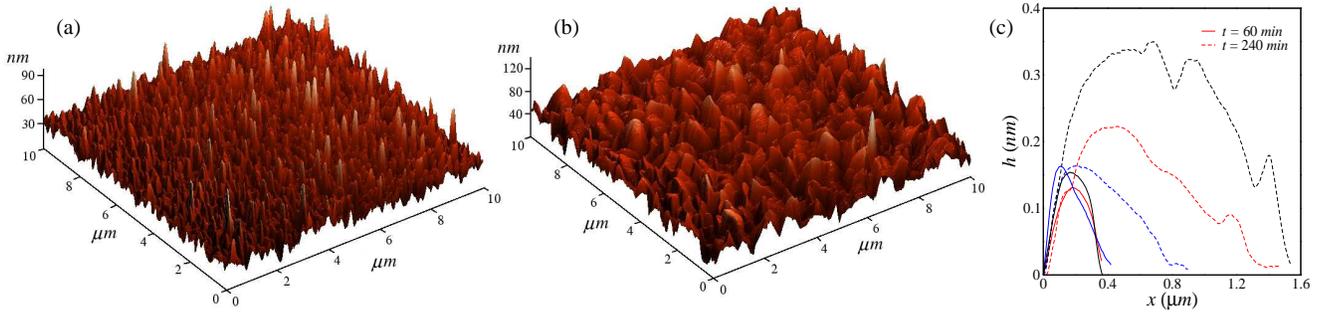}
\caption{(Color online) $10 \mu m$ $\times$ $10 \mu m$ AFM images for deposition times (a) $t=60 min$ and (b) $t=240 min$.
c) Typical grain shapes for $t=60$ (full lines) and $t=240 min$ (dashed lines).}
\label{fig1}
\end{figure*}

Details of the growth procedure of $CdTe$ films by hot wall epitaxy can be found in Ref. \cite{lowtemp}.
Prior to film growth, $Si(001)$ substrates were etched with a 2$\%$ HF solution to remove the native oxide layer.
The samples were grown from a single high purity $CdTe$ solid source, at substrate temperature of
$250\,^{\circ}\mathrm{C}$. Growth times range from 15 to $240 min$, with constant rate $2.2\mathring{A}$/s
for all samples, corresponding to thickness between $0.2\mu m$ and $3.2\mu m$. Surface characterization
was performed in air by AFM using a Ntegra Prima SPM in contact mode. Surface topographies
of 3 to 8 different regions of each sample were scanned, producing images of $10\mu m \times 10\mu m$ areas with
$1024 \times 1024$ pixels. This size was chosen so that morphological properties in domains smaller and
larger than the average mound size could be simultaneously investigated.

Figures \ref{fig1}a and \ref{fig1}b show typical surface morphologies of films grown for 60 and 240 minutes,
respectively. The mounded morphology typical of polycrystalline samples is observed, and the average mound size
increases in time. Indeed, it was already shown that $(111)$ grains grow faster than the ones
with other crystallographic orientations \cite{lowtemp}, which tend to be covered.
Height profiles of typical grains are shown in Fig. \ref{fig1}c to reveal mound coalescence and
change of aspect ratio. For short times, sharp isolated surface structures are found with size
$\lesssim 0.3 \mu m$,
which probably originate from islands formed in submonolayer regime. For long times, the mound basis
are 2-4 times larger (close to $1\mu m$), while their heights barely increase by a factor 2, and a multi-peaked shape is
observed, which indicates a coalescence mechanism.

\begin{figure}[!t]
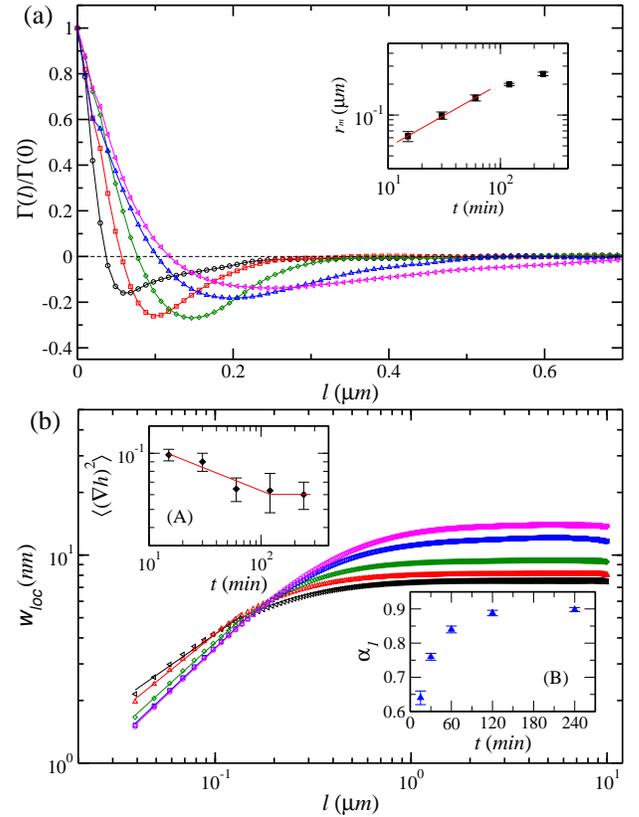

\includegraphics*[width=8.10cm]{Fig2a.eps}
\includegraphics*[width=8.10cm]{Fig2b.eps}
\caption{(Color online) a) Slope-slope correlation function $\Gamma(l)$ normalized to $\Gamma(0)$ against $l$.
The inset show the first minima of $\Gamma$ as functions of time.
b) Local roughness versus box length $l$, for times 15 (black triangles left), 30 (red triangles up), 60 (green diamonds), 
120 (blue squares) and 240 minutes (magenta circles). Full lines indicate the linear fits used to extract the exponent $\alpha_1$. 
The insets show the local squared surface slope (A) and the exponents $\alpha_1$ (B) against time.}
\label{fig2}
\end{figure}

The mound size is estimated from the first minima of the slope-slope correlation function
$\Gamma(l,t) \equiv \left\langle \nabla h(x+l,t) \nabla h(x,t) \right\rangle $ \cite{siniscalco},
shown in Fig. \ref{fig2}a for several growth times.
Those minima ($r_m$) are shown in the inset of Fig. \ref{fig2}a as a function of
growth time. For $t\lesssim 60 min$, before mound coalescence (see Fig. \ref{fig1}c),
it scales as $r_m\sim t^n$, with $n = 0.62(2)$.
The lateral correlation length $\xi$ is of the same order of the mound size in this regime, thus
$n$ is the inverse of the dynamic exponent, $z = 1/n = 1.61 (5)$, which is in excellent agreement with the
KPZ value \cite{Marinari}. For longer times, the inset of Fig. \ref{fig2}a shows $r_m<0.3\mu m$,
which is not the size of the multi-peaked mounds in Fig. \ref{fig1}c. Instead, the long-time estimates
of $r_m$ represent the size of the original
mounds that are constrained upon coalescence, providing a small coarsening exponent $n \approx 0.34$
in this regime. This value is similar to other systems dominated by surface diffusion \cite{etb},
but not representative of long lengthscale surface fluctuations.

\begin{figure}[!ht]
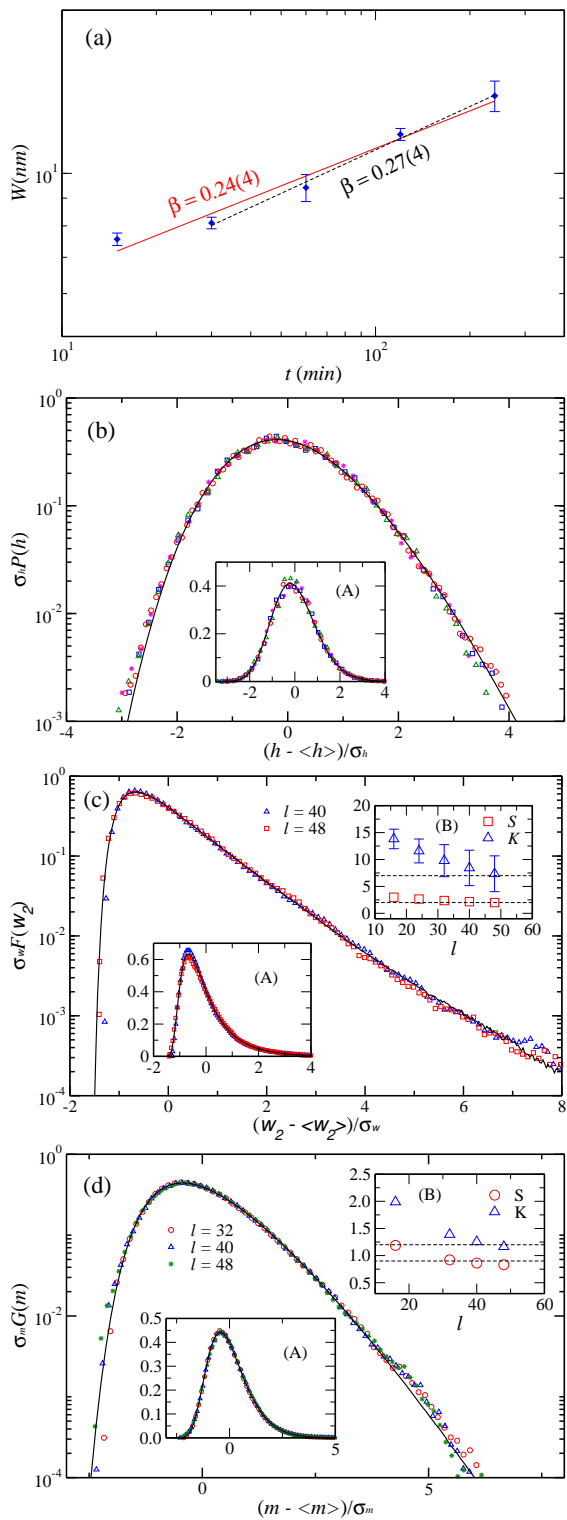

\includegraphics*[width=7.40cm]{Fig3a.eps}
\includegraphics*[width=7.40cm]{Fig3b.eps}
\includegraphics*[width=7.40cm]{Fig3c.eps}
\includegraphics*[width=7.40cm]{Fig3d.eps}
\caption{(Color online) a) Global surface roughness versus time. b) HDs for different AFM images.
c) Average SLRDs for different box sizes. d) Average MRHDs for different box sizes.
Insets (A) show the distributions in linear scale.
Insets (B) show the skewness and kurtosis of the distributions.
In  all cases $\sigma_{X} \equiv \sqrt{\left\langle X^2\right\rangle  - \left\langle X\right\rangle^2 }$.}
\label{fig3}
\end{figure}

The local roughness $w(l,t)$ of a sample is measured as the average height fluctuation in
boxes of lateral size $l$ that scan the film surface. Fig. \ref{fig2}b shows $w$ as a function of $l$
for several growth times. For short times and small sizes $l$,
$w$ measures the roughness of a single mound surface,
which evolves from sharp to rounded shape, respectively with larger and smaller height fluctuation.
This leads to the time decrease of $w$, corresponding to a decrease of the local slope,
which is shown in the inset (A) of Fig. \ref{fig2}b.
For comparison, this is the opposite of systems with anomalous roughening \cite{ramasco}.
In the first few minutes of deposition, island heights rapidly increase because substrate
wetting is not energetically favorable (a signature of the initial Volmer-Weber growth mode observed 
for this system \cite{suka}), but the inset (A) of Fig. \ref{fig2}b shows the subsequent
regime in which the substrate is fully covered and local slopes relax until saturating,
while long wavelength fluctuations ($w$ for large $l$) increase.

The inset (B) of Fig. \ref{fig2}b shows the initial slope $\alpha_1$ of the $\log{w}\times \log{l}$ curves,
which increases in time from $0.62$ ($t=5 min$) to $0.9$ ($t\geq 120 min$). This slope should not be interpreted as
a local roughness exponent, which characterizes long wavelength fluctuations. Instead, $\alpha_1$ is measured for
$l\sim 0.3\mu m$, which is smaller than or close to the average mound size \cite{tiago3}, thus it
characterizes the intragrain surface morphology. The estimates of $\alpha_1$ are fully consistent with
the values obtained in grainy surface models of Ref. \protect\cite{tiago3},
which vary from $\alpha_1\sim 0.7$ for sharply peaked grains (similar to the short time structures of Fig. \ref{fig1}c)
to $0.85\leq \alpha_1\leq 1$ for grains with flat or weakly rounded surfaces (similar to the long time
structures of Fig. \ref{fig1}c). Those models also explain the values of $\alpha_1$ reported
for growth of $CdTe$ on glass \cite{Nascimento} and on glass covered by fluorine-doped tin 
oxide \cite{Mata}.

Now we turn to the analysis of large lengthscale features and distributions, providing striking evidence of 
KPZ kinetics, which is related to the polycrystalline packing in the growing films.

Fig. \ref{fig3}a shows the time increase of the global roughness of the samples, $W$, which is expected to
scale as $W\sim t^\beta$, where $\beta$ is the growth exponent. A linear fit of the data
gives $\beta=0.24(4)$, which is in excellent agreement with the KPZ exponent $\beta \approx 0.24$ \cite{Marinari}.
The smallest time point in Fig. \ref{fig3}a slightly deviates from the linear fit, probably
due to a transient dynamics, commonly observed in KPZ systems \cite{tiago3}.
However, the growth exponent measured without that data point is
$\beta=0.27(4)$, which is still near the KPZ value.

In Fig. \ref{fig3}b, the HDs $[P(h)]$ of four $CdTe$ samples at the largest growth time are compared with the
Gumbel distribution with parameter $m=6$ \cite{gumbel}, which was recently shown to fit the
(numerically calculated) HD of KPZ in the growth regime of $2+1$ dimensions  \cite{SidTiaSil3}.
The excellent agreement in log-linear and linear-linear plots (see inset of Fig. \ref{fig3}b),
is the first experimental confirmation of this KPZ distribution in $2+1$ dimensions.
Small deviations between the experimental data and the theoretical curve are only observed for
the smallest heights ($h-\langle h\rangle<2\sigma_h$), indicating an increased density of
these deeper regions, which tend to appear between the surface grains. 

The visually observed collapse of distributions in Fig. \ref{fig3}b is reinforced by comparing the skewness
$S\equiv \left\langle X^{3} \right\rangle_{c}/\left\langle X^{2} \right\rangle_{c}^{3/2} $ and kurtosis 
$K\equiv \left\langle X^{4} \right\rangle_{c}/\left\langle X^{2} \right\rangle_{c}^{2} $
($\left\langle X^{n} \right\rangle_{c}$ is the $n^{th}$ cumulant of the fluctuating
variable $X$). For $CdTe$ films, we obtain $S=0.34(1)$, while numerical simulations of KPZ models give $S=0.42(1)$
\cite{healy,SidTiaSil3}, but this discrepancy is
expected from the enhancement of the experimental left tail, as explained above.
Furthermore, the numerical value was obtained 
in the asymptotic limit $[S (t\rightarrow \infty)]$ and, in general, smaller $S$ values are observed
in finite times \cite{SidTiaSil3}.
On the other hand, $K=0.3(1)$ for the $CdTe$ films is in good agreement with the KPZ value $K=0.34(2)$
\cite{healy,SidTiaSil3}.

Roughness distributions \cite{Foltin,Antal} are known to have much weaker finite-size corrections than
scaling exponents and HDs. Using digitalized images of a film surface, the suitable quantity to measure
is the squared local roughness distribution (SLRD) \cite{Antal} in boxes of size $l$ much larger than
the pixel size and much smaller than the whole image size, so that a large number of microscopic environments
can be sampled. Fig. \ref{fig3}c shows the SLRDs $[F(w_2)]$ of $CdTe$ films with two box sizes and of
KPZ models in $2+1$ dimensions \cite{FabioGR}, with an excellent data collapse in three orders of magnitude
of $F(w_2)$. Inset (A) of Fig. \ref{fig3}c confirms the agreement
near the SLRD peaks. Inset (B) of Fig. \ref{fig3}c confirms the convergence of the amplitude ratios $S$ and $K$
to the KPZ values as $l$ increases, i. e. approaching a continuous limit.
Also note the stretched exponential decay of SLRDs in Fig. \ref{fig3}c, which is the main signature of this KPZ
distribution in $2+1$ dimensions \cite{FabioGR} (in contrast to simple exponential decays of linear growth
equations and other Gaussian interface models \cite{Antal}).

Finally, maximal relative height distributions (MRHDs - $G(m)$) \cite{Raychaudhuri} of the $CdTe$ films
were also measured in several box sizes. They are compared in Fig. \ref{fig3}d with the universal MRHD
of the KPZ class in $2+1$ dimensions, again showing excellent agreement in two orders of magnitude.
Agreement in the peaks is confirmed in inset (A) and convergence of $S$ and $K$ to the KPZ values is shown in
inset (B) of Fig. \ref{fig3}d.

The importance of comparing distributions is clear if one observes that
the scaling exponents obtained here are close to those of diffusion-dominated growth governed by the
Villain-Lai-Das Sarma (VLDS) equation, which are $\beta\approx 0.20$, $\alpha\approx 0.67$, and $1/z\approx 0.30$.
Similar results in Refs. \cite{Mata,Nascimento} led to that proposal.
However, comparison of HDs, SLRDs, and MRHDs discard the VLDS interpretation; for 
instance, the SLRD of the VLDS class is Gaussian, in striking contrast to the stretched exponential of
$CdTe$ films. The simultaneous calculation of exponents and visual inspection of surface
features is also essential, since it led to the interpretation of $\alpha_1$ as representative of intragrain
features (not a Hurst exponent) and to the connection between the long time exponent $n$ and the restricted
growth of coalescing mounds. This establishes a reliable procedure to check the universality class of a
growing process and, consequently, to interpret its growth dynamics.

Thus, understanding the origin of KPZ scaling is in order.
During the growth of $CdTe$ films, the morphology of a crystalline grain
is not templated by the shapes of the previously formed grains. Instead, that morphology follows from
an interplay between intragrain surface energetics (that determine the favorable shape of that grain)
and the constraints imposed by the neighborhood. This leads to a complex packing of crystalline grains.
A simple illustration is provided by the models of Ref. \protect\cite{tiago3}, in which a new cubic grain is firmly
attached to the boundary of the grains below it, but does not fill all available space in their neighborhood
because its shape is constrained to be cubic. This aggregation mechanism has the same effect of the
lateral aggregation in ballistic aggregation \cite{barabasi}. It generates excess velocity,
which is the landmark of KPZ scaling.

In summary, a consistent analysis of the dynamic scaling properties of $CdTe$ films grown on $Si(001)$ substrates
is presented, giving the first experimental evidence of the universal KPZ distributions of heights,
maximal heights, and roughness in the growth regime in $2+1$ dimensions.
The evolution of mound shape determines the local roughness scaling
and the non-(111) mound coverage followed by coalescence of (111) mounds provides a scenario similar to 
typical KPZ models, explaining the excess velocity. Our results also shows that the analysis of distributions
is a powerful and essential tool in the study of universality classes of growth processes, which will
motivate future works in the field to go beyond the direct comparison of estimated scaling exponents and
model results.

\acknowledgments

The authors acknowledge the support from FAPEMIG, FAPERJ, CAPES and CNPq (Brazilian agencies).


\end{document}